\documentclass[prb,twocolumn]{revtex4}

\usepackage{graphicx}
\usepackage{dcolumn}
\usepackage{bm}

\begin{document}

\title{Modelling charge self-trapping in wide-gap dielectrics: \\
Localization problem in local density functionals}
\author{Jacob L. Gavartin, Peter V. Sushko, and Alexander L. Shluger}  
  
\affiliation{Department of Physics and Astronomy, University College London, 
Gower Street, London WC1E 6BT, UK} 
  
\date{\today} 


\begin{abstract}
We discuss the adiabatic self-trapping of small polarons within the density 
functional theory (DFT). In particular, we carried out plane-wave  
pseudo-potential calculations of the triplet exciton in NaCl and found
no energy minimum corresponding to the self-trapped exciton (STE)  
contrary to the experimental evidence and previous calculations. 
To explore the origin of this problem we modelled the self-trapped hole in 
NaCl using hybrid density functionals and an embedded cluster method. 
Calculations show that the stability of the self-trapped state of the hole
drastically depends on the amount of the exact exchange
in the density functional: at less than 30\% of the Hartree-Fock exchange, only 
delocalized hole is stable, at 50\% - both delocalized and self-trapped states 
are stable, while further increase of exact exchange results in only the 
self-trapped state being stable.
We argue that the main contributions to the self-trapping energy such as
the kinetic energy of the localizing charge, the chemical bond formation of 
the di-halogen quasi molecule, and the lattice polarization, are represented 
incorrectly within the Kohn-Sham (KS) based approaches. 
\end{abstract}

\maketitle

\section{Introduction}

Despite considerable progress in studies of self-trapped
excitons and polarons, the dynamics of early stages of self-trapping in specific 
systems is still poorly understood. The conceptual difficulty primarily lies 
with the fact that the quasi particle in question undergoes a transition 
between the free (delocalized) state and the localized one. In case of 
small polarons this transition is also associated with a substantial local 
lattice distortion.    
Within an adiabatic approximation one can describe self-trapping 
in terms of a potential energy surface (PES) connecting free and
localized states. Then a microscopic model of self-trapping will
involve a characterization of this PES, i.e. relevant atomic coordinates,
relative energies of the free and self-trapped states, the energy barrier
(if any) between them, as well as spectroscopic properties of free and
self-trapped species. 

Significant progress has been achieved in understanding of the 
conditions for polaron and exciton localization and self-trapping, as 
summarized in recent reviews 
\cite{Rashba_book,Song-Williams,Shluger-Stoneham,NIMB2000}. 
It has been suggested by Rashba \cite{Rashba} that in 
three dimensional dielectric crystals, the free and self-trapped forms of 
excitons may coexist, thus implying an energy barrier separating the two states. 
The height of this barrier affects the dynamics and characteristic time of 
self-trapping process. However, atomistic modelling has only been successful 
in calculating the structure and properties of strongly localized 
systems and never in modelling transitions between delocalized and 
localized states. One should note that the barrier for self-trapping, if it 
exists, is not likely to exceed a few tenths of an electron volt 
\cite{Song-Williams}. 
Therefore, its experimental verification and theoretical calculation is 
extremely challenging.

It has been anticipated that further development of quantum mechanical 
techniques, especially Density Functional Theory (DFT), will allow 
one to close this gap and achieve predictive modelling of self-trapping 
or defect induced trapping process. However, recent attempts to calculate even 
well established models of small radius polarons have failed unexpectedly. 
Calculations of the triplet 
self-trapped exciton in NaCl using plane-wave DFT in 
the Generalized Gradient Approximation (GGA) predicted no stable 
state  \cite{ICDIM2000} in direct contradiction with the experimental evidence 
\cite{Song-Williams} and previous Hartree-Fock calculations 
\cite{Puchin-Shluger,Shluger-Tanimura}. On the other hand, Perebeinos 
{\it et al.} \cite{Perebeinos} using plane-wave local spin density (LSDA) approach 
predicted the existence of a marginally stable STE in this system, 
albeit higher in energy than the free exciton state. 
Song {\it et al.} \cite{Song_Corrales,Song_Jonsson,Song_Kresse} applied plane 
wave DFT to the triplet 
exciton in $\alpha$-quartz and similarly found the free exciton state to be 
more stable than the localized one. 
The delocalized solution has also 
been found in LSDA calculations of the hole trapping at the $Li^0$ center 
in MgO \cite{Dovesi}. 
Pacchioni {\it et al.} \cite{Pacchioni} and L{\ae}gsgaard {\it et al.} \cite{Stokbro} 
considered the hole trapped at an Al impurity in silica. Both groups concluded that 
the predicted structure of this defect strongly depends on the density functional 
used in the calculations: local and GGA functionals predicted only the delocalized 
hole to be stable, again in contradiction with the experiment. Importantly, 
the non-local density functionals predict more 
localized states for this system. 

An apparent bias of DFT calculations towards the delocalized electronic states 
was attributed to the self-interaction error inherent in the local or semilocal
GGA type approximations, which are central to the Kohn-Sham method \cite{Parr-Yang}. 
It is unclear, however, to what extent the local approximation affects qualitative 
results (localized versus delocalized states), and what is the role of other factors 
in the calculation, such as boundary conditions, basis set completeness, and 
pseudopotential approximation. 

In this paper, we consider a triplet exciton and a hole in the
archetypal ionic insulator NaCl. These defects 
have been studied extensively both experimentally and theoretically 
\cite{Song-Williams,Puchin-Shluger,Testa} and thus 
provide good test systems. Previous calculations were 
carried out mainly in small embedded cluster models using the Hartree-Fock 
method and therefore were unable to treat delocalized states and take full 
account of the electron correlation. We would like to model a transition from 
delocalized to self-trapped exciton state and for this purpose use the 
plane-wave DFT method. 
We analyse the effect of size of a periodic supercell and the related question
of spurious multipole interactions in this system and conclude that no
stable self-trapped state for the exciton or a hole is predicted within the 
GGA DFT framework, once these factors are eliminated. To separate
the localization problem from the effects of periodic boundary conditions (PBC)
we then consider a hole within the embedded cluster approach and find that the 
non-local contribution to the exchange interaction is decisive in the description 
of self-trapped states.

The paper is organized as follows. In the next Section we give a brief 
account of the microscopic models of the self-trapped exciton and hole 
in alkali halides. Next we outline the details of the periodic 
plane-wave and embedded cluster procedures used. The results of calculations 
for an exciton and a hole are presented respectively in Sections IV and 
V, followed by discussion in section VI. 

\section{Self-trapped exciton and hole in alkali halides: background}

\begin{figure}
\includegraphics[width=\columnwidth]{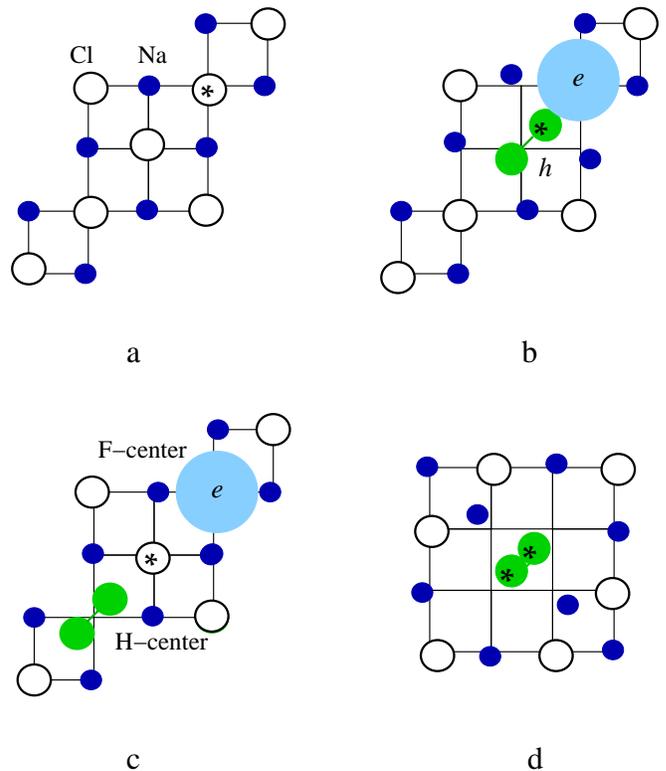}%
\caption{Schematic representations of the perfect rock-salt structure (a); of 
the triplet self-trapped exciton (b), the closest separated F-H pair (c), 
and the $V_k$-center (d) in an alkali 
halide crystal. The marked atoms were constrained during the optimization 
procedure.}
\label{fig1}
\end{figure}

Most of the alkali halides at normal pressure and temperature assume
the face centered cubic structure (Fig. ~\ref{fig1}a) (except cesium halides which 
are simple cubic).
Self-trapped excitons (STE) and holes in these crystals exhibit the features of both 
molecular and dielectric polaron character. In particular, the self-trapped hole 
(otherwise called a $V_k$-center) is known to be localized on two adjacent halogen 
ions forming a $X_2^-$ molecular ion (Fig. 
~\ref{fig1}d). There is no experimental evidence pointing to existence of a
barrier for the hole self-trapping in alkali halides 
\cite{Song-Williams,Aluker_book}, suggesting that the free hole is 
unstable. In particular, Lushchik \cite{Lushchik_book} {\it et al.} estimated that the 
mean free path of a free hole in NaCl before self-trapping does not exceed 30 
$a_0$ ($a_0$ is the shortest Na-Cl distance in the perfect lattice). The hole 
self-trapping energy, which is the difference between the energies of the fully 
delocalized and localized states, is larger than the activation energy for the 
$V_k$ center diffusion ($\sim 0.4$ eV in NaCl) \cite{Tanimura3}. Therefore, it is 
expected that the description of such a deep state is well within the reach of the
DFT theory. 

The model and the geometric structure of the $V_k$-center in alkali halides has 
been first proposed on the basis of the analysis of the electron spin resonance
data \cite{Castner,Schoemaker} 
and later refined in numerous theoretical calculations
\cite{Testa,Kantorovich_vk,Shluger_vk,Shluger_vk1}. The structure 
and stability of the $V_k$-center is primarily determined by the chemical bond 
formation between the two halogen ions, assisted by the lattice 
polarization.  

Self-trapped triplet excitons in alkali halides can be formed either directly by 
excitation in the exciton band or as a result of recombination of electron with a 
self-trapped hole, $V_k$-center. It is currently accepted that the STE in alkali 
halides consists of a hole localized on the $X_2^-$ molecular ion, and an electron 
localized in its vicinity. The so-called on-center and 
off-center configurations of triplet STE are considered in the literature
\cite{Song-Williams}. In the on-center configuration the $X_2^-$ molecular ion 
equally occupies two nearest anion sites (as in Fig. ~\ref{fig1}d), 
with the electron cloud symmetrically localized around it and has the $D_{2h}$ 
point 
symmetry group. It is therefore called a ($V_k$ + e) configuration. Williams and 
Song \cite{Leung} suggested that in some alkali halides  at least, the on-center 
configuration of the STE is unstable relative to an 
off-center displacement of the  $X_2^-$ ion along the $<110>$ crystallographic 
direction 
towards one of the anion sites (Fig. ~\ref{fig1}b). The center of mass of the 
electron 
component in this configuration is localized near the other (vacant) anion site, 
thus 
giving the $C_{2V}$ symmetry to the STE. This model of the triplet STE 
in alkali halides has been further developed using effective potential 
and embedded cluster {\it ab initio} Hartree-Fock calculations, as reviewed by 
Song and Williams \cite{Song-Williams}. 
The  STE in NaCl has also been studied using {\it ab 
initio} embedded cluster methods by Puchin \cite{Puchin-Shluger,Puchin_2} 
{\it et al.}, who confirmed the existence of the off-center STE configuration 
and demonstrated the importance of the electron correlation in 
determining the properties of the STE. 

Triplet STE created by irradiation may subsequently annihilate 
(radiatively or non-radiatively) restoring the perfect lattice, or decay into a 
Frenkel pair of lattice defects: 
an electron in an anion vacancy (F-center) and an interstitial halogen atom 
in the form of an $X_2^-$ molecular ion occupying one halogen site,  the H-center, 
(Fig. ~\ref{fig1}c). 
The energy barrier between the off-center STE and the closest F-H pair has been 
calculated in alkali chlorides and bromides by Song et al.
\cite{Song-Leung-Williams} and also recently in KBr by Shluger and Tanimura 
\cite{Shluger-Tanimura}. The dissociation barrier for the STE is found to be very 
low (0.07 eV for NaCl and 0.03 eV for KBr) \cite{Song-Leung-Williams}. 
The activation energy for the STE diffusion in NaCl is somewhat larger \cite{Tanimura3} 
$\sim 0.13$ eV.

At the outset of this work we hoped to use the plane wave periodic 
DFT methods to treat both delocalized and localized states and to study the 
adiabatic potential for self-trapping of triplet exciton in NaCl. The 
results presented below highlight the limited applicability of the existing 
functionals to the localization problem.

\section{Calculation procedure}

\subsection{Periodic DFT calculations}

In order to model the triplet exciton self-trapping we consider the adiabatic
coordinate corresponding to a displacement, $\Delta$, 
of one chlorine ion (hereafter designated as $Cl^*$) along the 
$<110>$ crystalline axis with all other surrounding atoms \cite{note1} relaxed at 
each position of the $Cl^*$. The displacement $\Delta$ is measured in units of 
$1/8$ of the plane diagonal ($a_0\sqrt{2}$). By virtue of construction, the 
specified adiabatic coordinate connects the free exciton state ($\Delta=0$) (Fig. 
\ref{fig1}a) with the STE configuration
 ($ \Delta \sim 4-5$) (Fig. \ref{fig1}b), and the closest F-H pair separated by one 
lattice anion, $\Delta \sim 8$) (Fig. \ref{fig1}c). The considered potential energy
surface (PES) allows for both on- and off-center forms of STE as well as for the 
transient one-centre excitons proposed by Shluger and Tanimura 
\cite{Shluger-Tanimura}. 
However, it should be emphasised, that an existence of the potential
energy barrier along this coordinate does not rule out a 
possibility of alternative self-trapping paths with still lower energy barriers.

The PES along the coordinate $\Delta$ was 
calculated using the plane-wave density functional approach implemented in the VASP 
code \cite{VASP1} with the GGA density functional due to Perdew and Wang \cite{PW91} 
and the ultra-soft pseudopotentials \cite{Vanderbilt} as supplied by Kresse and 
Hafner \cite{Pseudo2}. 
The self-consistent triplet excited state was modelled by constraining the 
total spin of the system to S=1, so as to obtain the lowest 
energy state of given multiplicity.

To study the effect of the super-cell size, the exciton PES was calculated in
unit cells containing 32, 64, 108 and 144 ions. The 
details of the super-cells are given in Table ~\ref{tab1} together with the sets 
of k-points used and number of atoms explicitly relaxed. 
At each position of $Cl^*$ the specified ions were relaxed, so that the maximum 
force acting on any individual atom did not exceed 0.04 $eV/\AA\ $ \cite{note1}.
Most of the calculations were performed with an energy cut-off of 285 eV. 

As shall be seen below, the calculated PES for the exciton self-trapping does
not display a minimum for the STE. Furthermore, the shape of PES strongly varies
with the size of the supercell. In order to separate
the effects of the boundary conditions and density functionals, we considered 
the formation of the $V_k$-center in NaCl using an 
embedded cluster method implemented in the GUESS code \cite{Sushko_2000a,GUESS2}.

\begin{table}
\caption{The supercells used in the exciton calculations, specified by the 
number of atoms, Cartesian translation vectors (given in units of the shortest 
Na-Cl distance $a_o$) and the number of irreducible k-points used in 
the calculations. The last column shows the value of the optimized lattice
constant $a_0$ for the ground and excited states.}
\begin{ruledtabular}
\begin{tabular}{lcccc}
Atoms  & Translation vectors ($a_0$) & No k-points & 
\multicolumn{2}{c}{$a_0$ \AA\,} \\
(relaxed) \protect\cite{note1} &  &    & s=0   & s=1    \\
\hline
2        & (1 1 0), (1 0 1), (0 1 1)  & 35 & 2.845 & -      \\
32 (26)  & (4 0 0), (0 4 0), (2 2 2)  & 4  & 2.83  & 2.87   \\
64 (48)  & (4 0 0), (0 4 0), (0 0 4)  & 4  & 2.84  & 2.85   \\
108 (28) & (3 3 0), (3 0 3), (0 6 6)  & 3  & 2.83  & 2.84   \\
144 (46) & (6 0 0), (0 6 0), (0 0 4)  & 1  & 2.84  & 2.84   \\
\end{tabular}
\end{ruledtabular}
\label{tab1}
\end{table}

\subsection{Embedded cluster calculations}

In this method, a crystal is 
represented by a large finite nano-cluster, that is divided into several 
regions. Spherical region I at the centre of the 
nano-cluster includes a quantum-mechanically treated cluster (QC)
surrounded by interface ions and a region of classical shell model ions 
\cite{Dick_1958}. The remaining part of the nano-cluster is represented by 
classical non-polarizable ions.  All quantum mechanical, interface and 
classical ions (both cores and shells) in region I are allowed to relax 
simultaneously in the course of geometry optimization. Ions outside region I 
remain fixed and provide an accurate electrostatic potential within region I.
This approach allows one to take into account the defect-induced lattice 
polarization of a crystal region containing a few hundred atoms. An account of 
lattice polarisation outside region I can be extended to infinity 
using a polarizable continuum model and the Mott-Littleton 
approach \cite{Catlow-Stoneham}. In this approximation the polarization energy is 
proportional to the square of the excess charge in the lattice. In the case 
of the $V_k$-center, the charge remains constant, so the Mott-Littleton 
contribution cancels out in any energy differences. Therefore this term was 
neglected in the current calculations. 

The described scheme is implemented in the GUESS code \cite{Sushko_2000a,GUESS2}, 
which provides an interface between the GAUSSIAN package \cite{Gaussian_98A1} for 
calculation the electronic structure of the QC and a code providing the shell model 
representation of the rest of the crystal.  The total energy and the electronic 
structure of the QC is calculated by solving standard Hartree-Fock or
Kohn-Sham equations which include the matrix elements of the electrostatic potential 
due to all classical point charges in regions I and II computed on the basis 
functions of the cluster. Further details for the total energy evaluation in this
approach are given elsewhere \cite{Sushko_2000a}. 

\begin{figure}
\includegraphics[width=\columnwidth]{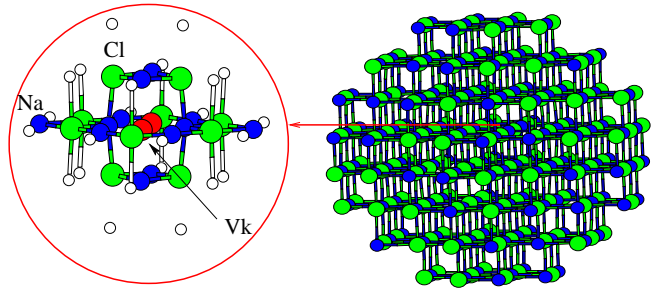}%
\caption{Illustration of the embedded cluster model used in the $V_k$-center 
calculation (only region 1 is shown). The quantum cluster is shown enlarged. The 
small white balls represent the sodium interface atoms}
\label{fig2}
\end{figure}

A cubic nano-cluster containing 8000 (20x20x20) ions was used in these 
calculations. Spherical region I of radius 11.5 \AA\, included 300 ions and a 
stoichiometric QC comprising 22 ions (Fig. ~\ref{fig2}). 

In our scheme, the charge density in QC interacts with the classical atoms
only electrostatically, which results in artificial polarization by the
classical cations nearest to the QC \cite{Ferrari_1997}. To avoid this, an 
interface region between the QC and classical ions in region I was introduced. 
In the present model it includes 20 Na ions represented by effective core 
pseudopotentials \cite{Wadt_1985}, and a single $s$-type basis function 
(see Fig. \ref{fig2}). Therefore all QC anions are fully coordinated by either 
quantum or interface ions.  For the quantum cluster we have used an all electron 
Gaussian type atomic basis \cite{crystal-basis} 
optimized for the NaCl crystal ($8-511G$ for Na and $86-311G$ for Cl).

The shell model ions in region I interact via pair potentials due to Catlow {\it et al.}
\cite{NaCl-pot},  which were slightly modified to represent more accurately the
static dielectric constant.  Modified potentials give a 
static equilibrium $Na-Cl$ distance of $2.79 \AA\ $ and fairly well reproduce the 
elastic, dielectric and phonon properties of the perfect NaCl crystal. 

The embedded cluster calculations were carried out using three different 
density functionals:
the Hartree-Fock, Becke hybrid three parameter (B3), and so-called half-and-half 
functionals (BH\&H) \cite{B3LYP}. The latter two incorporate 20\% and 
50\% of the Hartree-Fock exchange, respectively \cite{B3LYP}. The hybrid 
exchange functionals were used in conjunction with the Lee, Yang and 
Parr correlation functional (LYP) \cite{LYP}.  

Consistency of the interactions between the classical and quantum 
regions was tested by calculating an ideal QC embedded into an ideal
classical lattice. The optimized $Na-Cl$ distances inside the
QC and those at the cluster boundary differ from the $Na-Cl$ distances 
in the rest of the nano-cluster by $ca.$ 1\% and 3.5\% respectively.

\section{Exciton self-trapping in NaCl}

Exciton self-trapping is associated with a strong lattice relaxation.
It has been suggested that this relaxation is associated with the
significant volume change \cite{Tanimura-Tanaka}. To account for this effect,
we optimized the lattice constant of the supercells in both the ground 
and excited triplet states using the VASP code. The results are presented in (Table 
{\ref{tab1}). The calculated ground state equilibrium lattice constant $a_0=2.84$ 
\AA\ is $\sim 2\%$ larger than the experimental value, extrapolated to 0 K,
of 2.79 \AA\ . We also observe a slight 
increase of the perfect lattice constant in the triplet excited state in smaller 
supercells.

Next, we calculate the  potential energy surface for the triplet exciton by displacing
a selected $Cl^*$ ion along the $<110>$ crystallographic axis and relaxing 
all the surrounding atoms at each $Cl^*$ position (Fig. \ref{fig1}a,b,c). The PES for 
the triplet state calculated in different supercells is shown in Figure \ref{fig3}. 
Additional information on the lattice relaxation and spin density 
changes along the $Cl^*$ displacement coordinate can be found on our web page 
\cite{my_web}. It is seen that the obtained PES varies significantly with 
the size and shape of the supercell, in which it was calculated. 
In all calculations the free exciton configuration is predicted to be the lowest 
energy state, as reported previously \cite{ICDIM2000,Perebeinos}. 
The off-center STE configuration corresponds approximately to 
($\Delta \sim 5$). This state is predicted to be marginally stable only in the 
smallest supercell used (identical to the one used by Perebeinos {\it et al.} 
\cite{Perebeinos}). We also observe that the energy difference between the free 
exciton and the expected self-trapped state increases with the size of the 
supercell. 

\begin{figure}
\includegraphics[width=\columnwidth]{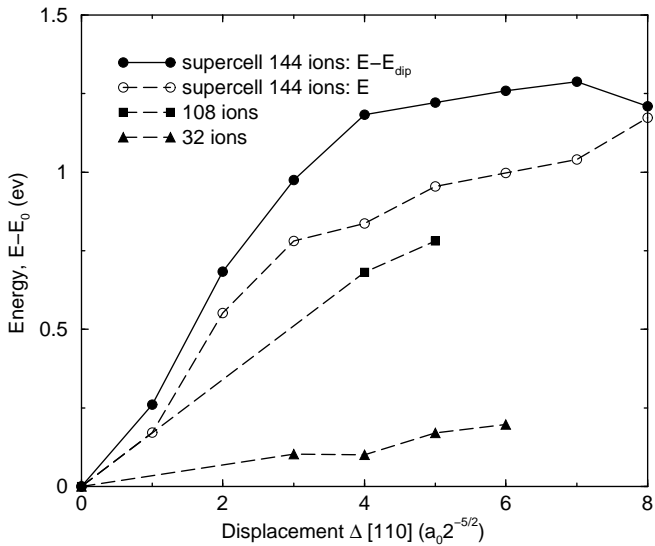}%
\caption{The potential energy surface for the triplet exciton as described in 
the text.
The energies are plotted with respect to the energies of the triplet state of a 
perfect
lattice (zero coordinate). $\Delta \sim 5$ corresponds to the expected minimum 
for
the STE (Fig. \protect\ref{fig1}b), $\Delta = 8$ corresponds to the next 
nearest-neighbour F-H pair (Fig. \protect\ref{fig1}c). The dipole corrected (solid line) 
and 
uncorrected (dashed line) energies are shown for the supercell of 144 ions.}
\label{fig3}
\end{figure}

We have singled out most probable causes of the 
significant dependence of the calculated PES on the size and shape of different 
supercells: i) defect volume change; ii) spurious electrostatic interaction between 
the supercells; iii) basis set completeness, and iv) pseudopotential approximation.
In the rest of this Section we discuss these aspects.  

To mimic the effect of the volume change associated with the self-trapping, 
we optimized the volume of the 144 ion supercell at each point of the adiabatic 
surface. The equilibrium lattice constant, $a_0$ has increased only slightly: 
from 2.840 \AA\ for a perfect
lattice ($\Delta=0$) to 2.848 \AA\ at the expected STE state ($\Delta=5$). 
The total energy of the STE has lowered by $\sim 0.1$ eV and the energy minimum 
for the STE did not exist.

As can be seen in (Fig. \ref{fig1}b), the charge distribution in off-center STE 
is associated with a substantial dipole moment. The strong dependence of the PES on the 
supercell size suggests that a significant contribution to the total energy can 
arise from the interaction between periodic dipoles persisting even for 
comparatively large supercells. The dipole-dipole interaction energy resulting 
from the super-lattice of dipoles can be evaluated according to the expression 
{\cite{Kantorovich_dip}:
\begin{eqnarray}
E_{dip} = - \frac{2G^3}{3\sqrt{\pi}}{\bf P}^2 + 
\frac{1}{2} P_{\alpha} T_{\alpha \beta} P_{\beta}, \label{eqn1.4.1}
\end{eqnarray}
where Einstein notation is adopted for the sums over repeated indices, 
$P_{i}$ denotes the Cartesian component of the dipole vector
${\bf P}$, $G$ is the Ewald parameter, and  
\begin{eqnarray}
T_{\alpha \beta} = \frac{4\pi}{v_c} \sum_{\textstyle g \neq 0}
\frac{{R_{g \alpha}R_{g \beta}}}{{R}_g^2}
e^{-{R}_g^2/4G^2} - G^3 \sum_{\textstyle i \neq 0} H_{\alpha \beta}(G{\bf R}_i).
\label{eqn1.4.2}
\end{eqnarray}
Here the first sum is over the reciprocal lattice vectors 
${\bf R}_g$ and the second - over the direct lattice vectors 
${\bf R}_i$ with the zero vector ($i=0$) being excluded to avoid 
self-interaction. In addition,
$H_{\alpha \beta}({\bf y})= - \delta_{\alpha \beta}H(y) +
y_{\alpha} y_{\beta}/y^2 [3H(y)+4/\sqrt{\pi} e^{-y^2}]$ and
$ H(y) = 2\sqrt{\pi} y^{-2}e^{-y^2}+erfc(y)/y^3$.
Equation (\ref{eqn1.4.1}) is a generalization of the Makov and Payne formula 
\cite{Makov-Payne} for the non-cubic supercells. 

To quantify the dipole-dipole interaction energy 
the dipole moment of the supercell needs to be calculated. For a neutral 
periodic system this must be defined as an invariant of a coordinate system \cite{Martin}:
\begin{eqnarray}
{\bf P}({\bf r}) = \Omega^{-1} \int_{cell} {\bf r} \rho ({\bf r}) d^3 {\bf r} +
\Omega^{-1} \int_{surface}  {\bf r} \left[ {\bf n \Pi }({\bf r}) \right]d{\bf 
s},
\label{eqn1.4.3}
\end{eqnarray}
where the first integral is taken over the volume of the supercell, and the 
second 
over the supercell surface, $\Omega$ is the volume of the supercell, 
$\rho({\bf r})$ 
is a charge density (including both electronic and atomic cores), ${\bf n}$ is 
an outward 
surface-normal unit vector, and ${\bf \Pi }$ is a local polarization defined via 
the equation:
\begin{eqnarray}
\nabla {\bf \Pi }({\bf r}) = -\rho ({\bf r}).
\end{eqnarray}
It should be noted, that within the PBC only the whole sum in 
Eq. ~\ref{eqn1.4.3} is an invariant with respect to the coordinate origin. 
However, the second integral there can be made small by an appropriate choice 
of a coordinate system. 
Then and only then the dipole moment of the supercell can be correctly 
estimated by the 
first integral in Eq. (\ref{eqn1.4.3}). To minimize the contribution of the 
surface integral we chose the supercell as a Wigner-Seitz cell,
where the origin of the charge density grid is located not on an atomic site, 
but on a volume interstitial position in the face centered cubic lattice.

The Cartesian components of the calculated dipole moment for each atomic 
configuration of the exciton along the adiabatic coordinate, $\Delta$, in the 
supercell of 144 atoms are shown in Fig. \ref{fig4}. The dipole moment varies 
non-monotonically along the $\Delta$ coordinate reaching a maximum near the 
expected STE configuration. Thus, the PES corrected for the 
dipole-dipole interaction may in principle contain a local minimum for the STE. 
To explore this possibility we calculated the value of the dipole energy 
(Eq. \ref{eqn1.4.1}) in different unit cells for the expected STE configuration  
($\Delta = 5$). For 
that we assumed that calculations in different unit cells yield approximately 
the same charge density for the STE and used the value for the dipole moment, 
{\bf P}, calculated in the 144 supercell for $\Delta=5$ (Fig. {\ref{fig4}). 
The obtained values of $E_{dip}$ are equal to  -4.85, -2.64, 0.18, and -0.27 eV 
for the 32, 64, 108, and 144 ion supercells, respectively, and are negative except 
for the 108 ion supercell. This difference is caused by the fact that this 
supercell has a rhombohedral shape whereas the other cells are rectangular.

\begin{figure}
\includegraphics[width=\columnwidth]{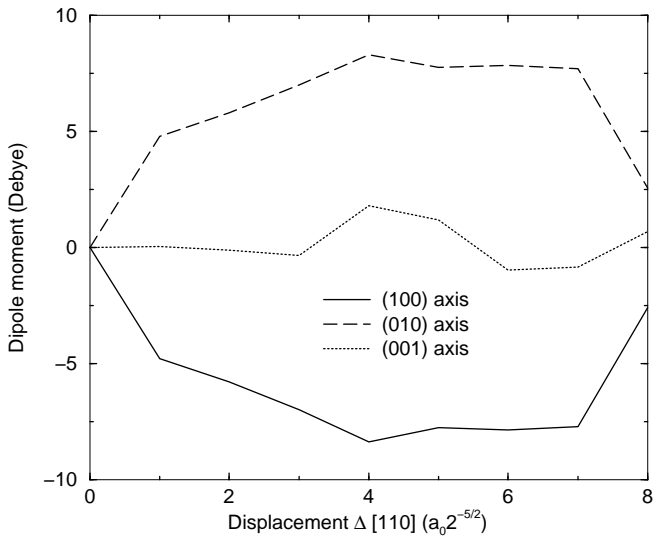}%
\caption{The electric dipole moment as a function of the adiabatic coordinate
$\Delta$, remaining details as in Fig. \protect\ref{fig3}}
\label{fig4}
\end{figure}

From the tendency shown in Fig. ~\ref{fig4}, we expect that the 
$E_{dip}(\Delta=5)$ values given above represent the largest corrections 
for different supercells. When 
subtracted from the total energy, $E(\Delta)- E_{dip}(\Delta)$, they in most 
cases enhance 
even further the difference between the free and off-center STE energies. The 
full PES corrected by the dipole-dipole contribution, $E(\Delta)- 
E_{dip}(\Delta)$, was calculated only for the largest 144 ion supercell where 
the correction is the smallest (see Fig. ~\ref{fig3}). Although the shape of the 
PES changes slightly, it does not show a minimum for any displacement $\Delta$. 
It is also clear that no energy minimum for the STE will occur in the 32 ion 
supercell after the dipole correction is taken into account. Our calculations 
clearly demonstrate that spurious electrostatic interaction in small unit 
cells may result in an artificial energy minimum for the STE. We suggest that the 
marginally stable STE configuration in the 32 atoms supercell 
reported by Perebeinos \cite{Perebeinos} {\it et al.} and also found in our calculation 
is precisely of this origin.

We should note that the $E_{dip}$ calculated using 
Eq. ({\ref{eqn1.4.1}) overestimates the electrostatic interaction between the 
supercells in PBC due to various approximations in Eq. ({\ref{eqn1.4.1}) :
  
1. The point dipole approximation assumed in Eq. ({\ref{eqn1.4.1}), breaks 
down whenever the dipole length is comparable with the supercell size. 
(Hence unrealistically large dipole energies are obtained in small 
supercells). 

2. Eq. ({\ref{eqn1.4.1}) does not account for a dielectric screening. 

3. The higher multipole contributions (dipole-quadrupole etc.) would 
generally have the sign opposite to $E_{dip}$ and may reduce the overall 
effect of spurious multipole interactions. 

4. Due to a strong multipole interaction in smaller supercells, the charge 
distributions, 
and therefore the dipole moments, predicted in different supercells may differ 
significantly.

In spite of these remarks we believe that further refinement of the calculations will
not affect the main conclusion - no stable state for the STE is predicted
by the GGA DFT approach. 

The reasons for the absence of a STE energy minimum in the local DFT 
approximation are not obvious. Several factors could result in the bias towards 
the delocalized solution. These include, for example, the choice of energy cut-off and 
k-point sampling. To check these issues, we compared calculations with cut-off 
energies of 220 eV and 285 eV and observed no significant difference in the STE 
relaxation. The k-point sampling is likely to affect more the energies of 
delocalized states. In fact we do observe a slight increase of the free exciton 
energy with the number of irreducible k-points. This reduces the 
energy difference between the free and the self-trapped exciton, though the
minimum for the STE still does not arise.
Another uncontrolled factor in the calculation is the pseudopotential 
approximation for core electrons. In particular, we employed a large core 
($1s^{2}2s^{2}p^{6}$) pseudopotential for sodium atoms and the polarization of 
cations was virtually neglected in our calculations. To examine the role of cation
polarization, we carried out the STE calculations with the Helium core pseudopotential 
for Na atoms for selected configurations in the 64 atoms supercell. 
This also did not affect our qualitative conclusions - no energy minimum was 
obtained for the STE. 

Our analysis suggests that the problem of localization lies not with
the technical details of the calculations but rather more fundamentally with 
the approximations involved in the GGA density functional. 
In particular, essentially uncontrolled self-interaction error in the local 
density approximation is the most likely cause. However, a detailed
characterisation of the involved energy terms is difficult for the exciton,
which comprises two strongly interacting spins with the energies near the 
top of the valence band and the bottom of the conduction band respectively.
It must be emphasised at this point that holes in alkali halides do self-trap 
in the form of $V_k$-centres, while the electrons do not.  
Hence, the hole self-trapping is also fundamental in the exciton localization.  
At the same time, self-trapping of the hole is a simpler problem in
that it involves localization of a single spin. However, being charged,
a hole presents some difficulty for studying within the PBC. In particular,
we obtained no self-trapping of the hole within plane wave DFT. 
Therefore, we resorted to modelling the hole within the embedded cluster 
method, that has the following advantages:
1. The atomic-type basis set ensures no bias towards the delocalized 
states;
2. The lattice polarisation effects paramount for charged polar systems 
are fully accounted for;
3. The use of the atomic basis sets allows the straightforward application
of a number of local and non-local density functionals, so that the properties of 
various functionals can be studied.  

\section{Hole trapping in NaCl}

The calculations of the hole (a total charge of the system is equal to +1) were performed 
within the embedded cluster model discussed in Section IIIB (Fig. 2).  
We calculated the total energy of the system as a function of the distance 
between the two Cl ions equally displaced along the $<110>$ axis towards 
each other from their respective perfect lattice sites (see Fig. ~\ref{fig1}d). 
All other ions in region I were relaxed simultaneously to their equilibrium 
positions for each Cl-Cl separation.

The potential energy surfaces calculated with different density functionals are 
depicted in Fig. \ref{fig5}. The zero energy in the top panel 
corresponds to the singly ionized perfect lattice calculated within each 
method. We observe that the shape of PES qualitatively depends on the amount of 
the Hartree-Fock exchange included in the density functional. The
B3LYP functional predicts one global minimum corresponding to a free hole, the 
BH\&H functional predicts two minima separated by a marginal barrier, while
the Hartree-Fock method gives a single energy minimum corresponding to the 
self-trapped hole ($V_k$-center) (top panel of Fig. \ref{fig5}). Furthermore, the 
energy difference between the free and self-trapped states ranges between -0.2 eV 
for B3LYP and 1.5 eV for the HF.

\begin{figure}
\includegraphics[width=\columnwidth]{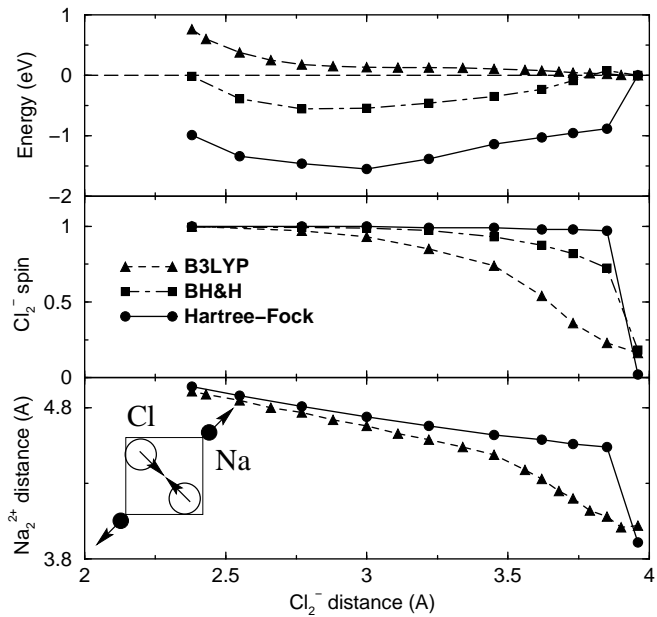}%
\caption{The adiabatic surface for the $V_k$-center (top panel), 
calculated with various density functionals. At each given $Cl_2^-$ distance,
all other atoms in the system were relaxed to the minimum energy geometry. 
Zero energy corresponds to the perfect crystal configuration (the extended 
hole).
The middle panel shows the dependence of the Mulliken spin population on 
the $Cl_2^-$ molecule as a function of the same adiabatic coordinate. The bottom 
panel shows the relaxation of the pair of Na atoms adjacent to the $V_k$-
center.}
\label{fig5}
\end{figure}

Different functionals also predict distinctly different  
spin localization. The analysis of the Mulliken 
spin population on the $Cl_2^-$ molecule (middle panel in Fig. \ref{fig5}) 
reveals that the extent of spin localization increases faster with decreasing 
$Cl_2^-$ distance when the amount of exact exchange is larger. 
In particular, the Hartree-Fock calculations predict the spin almost 
fully localized on $Cl_2^-$ at a mere 0.05 \AA\ displacement 
of two Cl ions from their respective lattice sites. The B3LYP 
functional, on the contrary, predicts very gradual spin localization with the 
decrease of the Cl-Cl distance. The variance in localization of spin in
different functionals also affects the lattice relaxation.
The bottom panel in Figure \ref{fig5} shows the distance between the two 
relaxed Na ions adjacent to the $Cl_2^-$ molecule as a function of the $Cl_2^-$ 
distance. It is seen that weak spin localization predicted in the B3LYP functional 
causes much smaller cation relaxation than that obtained in the Hartree-Fock 
method. 

\section{Discussion}

Our results highlight severe problems with applying
density functional theory to the self-trapping problem. The LSDA and GGA density 
functionals yield solutions biased towards the delocalized states to such an 
extent that no stable self-trapped state is predicted for either a hole or an 
exciton, contrary to experimental evidence. 

Let us emphasise that the degree of localisation of a polaron is closely linked to the 
properties of the single particle spectrum for the problem. For instance, spin density 
of the $V_k$-center is represented almost entirely by the single-particle density of the 
last occupied orbital in a majority spin, $\psi_s({\bf r}')$. The degree of localization 
of $\psi_s({\bf r}')$ is related to its energy, $\epsilon_s$, or more precisely, 
to the energy split of the $\epsilon_s$ from the corresponding band edge. In 
the KS method, $\epsilon_s$ is defined as \cite{Parr-Yang}:
\begin{eqnarray}
\epsilon_s = <\psi_s|-\frac{1}{2}\nabla^2 + v_{eff}({\bf r}) |\psi_s >,
\label{d1}
\end{eqnarray}
where
\begin{eqnarray}
v_{eff}({\bf r}) = v({\bf r}) +\int \frac {\rho({\bf r}')}{|{\bf r}-{\bf r}'|}d{\bf r}'
+v_{xc}({\bf r}).
\label{d2}
\end{eqnarray}
Here, $v({\bf r})$ represents the external potential of the atomic cores, the second term
is the classical electrostatic potential due to electrons of density $\rho({\bf r})$, 
(including the self-interaction), and $v_{xc}({\bf r})$ is a local exchange-correlation 
potential, where the self-interaction energy is partially cancelled.
How do the terms in Eq. (~\ref{d1}) contribute into self-trapping? 
One of the main factors facilitating self-trapping is lattice 
polarization. In alkali halide crystals, the second (and perhaps dominant) 
negative term is the energy gain due to the formation of the $\sigma$-bond (localised
doubly occupied orbital) in the $X_2^-$ molecular ion (X = F, Cl, Br, I). 
The factors favouring self-trapping 
are partially counterbalanced by an increase in the kinetic energy of 
localizing electrons. The latter results in a splitting of corresponding single 
particle levels from the edge of the valence band. 
First, we note that any perturbation in the external potential, $v({\bf r})$, alone 
cannot cause self-trapping, since it does not depend on the number of electrons or their 
state, and self-trapping does not occur in a neutral crystal in its ground state.
This is in contrast to polaron localization by defects, where the $v({\bf r})$ 
contribution to localization is dominant. 
Therefore the critical localization terms for self-trapping are contained in the 
Hartree (electronic polarisation) and exchange-correlation (bond formation) terms.

Apparently, an accurate calculation of these contributions presents a 
serious problem for the KS DFT, for the following reasons:

1) First, let us consider the formation of the $X_2^-$ quasi-molecule. 
For that purpose we calculate the dissociation curve for the 
free $Cl_2^-$ molecular ion using different density functionals and 
the Gaussian-98 package \cite{Gaussian_98A1}. 
We shall use as a reference the energy curve obtained by the coupled cluster 
method with double substitutions (CCD). In Fig \ref{fig6}, we observe that both 
the Unrestricted Hartree-Fock and all studied DFT functionals do not predict 
thed correct dissociation behaviour for this open shell ion. 
Moreover, local as well as hybrid density functionals (but not the Hartree-Fock 
method) violate the energy variational principle in that the predicted dissociation 
curves asymptotically 
tend to energies significantly lower than the corresponding dissociation limit 
(i.e. the sum of the ground state energies of the isolated negative chlorine ion 
and the neutral chlorine atom). The implication of this drawback for the self-
trapping is clear 
- the energy gain associated with a decrease of the inter-ionic distance in the 
$Cl^-_2$ from $\sim 4$ \AA\, (as in a perfect lattice) to the $V_k$-center 
equilibrium configuration 
($\sim 3$ \AA\,), is substantially underestimated by the DFT and 
overestimated by the UHF method.

\begin{figure}
\includegraphics[width=\columnwidth]{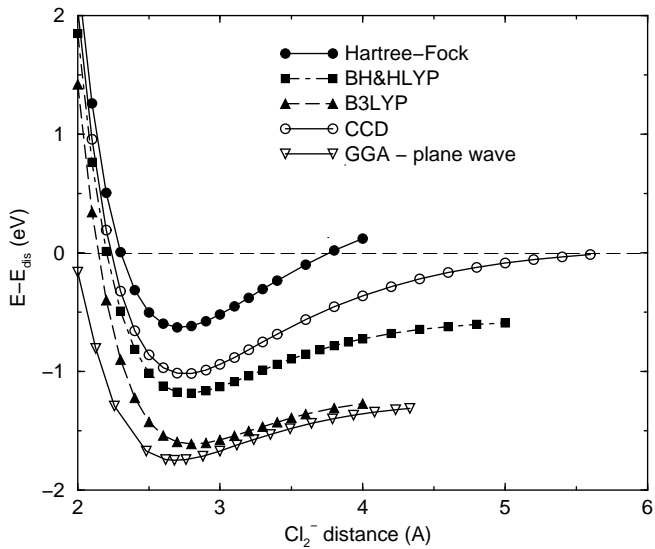}%
\caption{The dissociation curve  for the $Cl^-_2$ molecular ion calculated
using various density functionals ($E_{dis} = E(Cl^-)+ E_{Cl^0}$). All the local 
basis calculations are done using the G86-311 Gaussian basis set 
\protect\cite{crystal-basis}. 
In the plane wave GGA calculations the energy cut-off 285.3 eV is used. 
}
\label{fig6}
\end{figure}

Incorrect dissociation behaviour of radical ions in DFT
was first reported by Bally and Sastry \cite{Bally-Sastry}. The subsequent 
comprehensive analysis of the $H_2^+$ molecular ion by Zhang and Yang 
\cite{Zhang-Yang} revealed that the self-interaction error in local density 
functional for such systems becomes uncontrollably large.  
It can be demonstrated that the self-interaction error, associated with the single 
particle states, is positive and increases with their localization. Therefore, 
self-interaction, being minimized during self-consistency, will also lead to the 
less localized single particle states.

2) The polarization properties of a many electron system in the density 
functional 
theory are related to the exchange-correlation kernel \cite{Godby_1} 
$K_{xc}({\bf r,r}')=\frac{\delta^2E_{xc}}{\delta n({\bf r}) \delta n({\bf r'})}$. 
The exact exchange correlation kernel is shown to possess the 
so-called ultra-non-locality property, i.e. the diagonal element of its Fourier 
transform $K({\bf q},{\bf q})$ possesses a $O(1/q^2)$ divergence at small 
wavenumbers $q$. However, as first demonstrated in refs. \cite{Godby_1,Godby_2}, both 
LDA and GGA lead to $K_{xc}({\bf r r}')$ which does not have the required $O(1/q^2)$ 
divergence. One of the implications of the non-divergent
exchange-correlation kernel in KS DFT is that the macroscopic polarizability 
of the system (calculated as a KS polarizability) depends on the number of 
electrons and on the occupancy of the KS bands. That is, a calculated KS 
polarizability of the dielectric will be generally different in a singlet 
and triplet state, and for a system with or without the hole 
(both systems are technically metallic). Thus, the reliability of polarization 
contribution in KS DFT for the triplet or charged systems is also essentially 
uncontrolled. 

3) As mentioned earlier, the largely single particle character of the self-trapping 
problem implies that the total spin density may be represented by a single particle 
density of just one orbital. It is known however, that KS orbitals do not 
necessarily carry this physical meaning. In particular, 
by virtue of construction, the KS formalism yields the single particle 
eigenstates which minimize the kinetic energy (as calculated on single 
particle orbitals) \cite{Parr-Yang}. Thus, the kinetic energy contribution 
is always underestimated, leading generally to the least localized set of orbitals
for a given density. Another delocalizing factor is the self-interaction error in 
the Hartree term, which is only partially cancelled in the local approximations to 
the exchange-correlation potentials. Apparently, a positive self-interaction grows 
uncontrollably as the state becomes more and more localized. Therefore, any 
self-consistent procedure minimizing the total energy, also minimizes self-interaction 
by delocalizing the orbitals. It should be noted, that the effects of minimization of 
the kinetic energy and self-interaction in KS method have the same effect and cannot 
be separated, since any form of non-local self-interaction correction leads to the 
single particle equations distinct from those of KS.

Conceivably the depth of the self-trapping potential in  
polar dielectrics is unlikely to exceed 1 eV. It seems, that the 
spurious delocalizing contributions inherent in the KS method always outgrow the 
localizing potential, so the self-trapped state cannot be predicted in this
approach.

This problem is partially resolved by use of hybrid density functionals.
We have demonstrated for the holes 
that the self-trapped solution gradually becomes energetically more stable as the 
amount of the exact exchange is increased. However, 
the conventional hybrid functionals, such as B3LYP, failed to predict a stable
$V_k$-center, let alone its localization energy. This raises the question of
parametrisation strategies for semiempirical functionals, which are not clear
in the context of self-trapping.

Our calculations also highlighted another important methodological issue related 
to the calculations of the PES for polar systems within the periodic 
boundary conditions. The artificial multipole electrostatic interactions between 
supercells may drastically affect the shape of the PES, if the leading multipoles 
of the charge density vary along the chosen adiabatic coordinate. The dipole-dipole 
interaction in the triplet exciton modelled in a small supercell resulted in the 
appearance of the energy minimum for the STE configuration reported by Perebeinos 
{\it et al.} \cite{Perebeinos}. Interestingly, the authors of ref. \cite{Perebeinos} 
reported that no stable configuration could be found for the $V_k$-center. 
These results can be understood by recalling that in the $V_k$-center calculations 
the leading term in the electrostatic interaction between the supercells
is the monopole-monopole one. This term does not vary with the adiabatic 
coordinate and therefore does not affect the shape of the PES. 

Although artificial delocalization in KS based DFT theory can be expected, 
it is difficult to assess {\it a priori} when a specific functional will fail
in practical calculations. We suggest, that the problem with the localized states
in KS method will occur in all cases where the localization potential is provided
mainly by the electron-electron term. These include self-trapping phenomena, 
Jahn-Teller systems, low dimensional systems (e.g. Pierls instability), and 
systems with weak impurity perturbations. A list (by no means complete) 
of problematic DFT calculations where an artificial delocalization was reported 
includes: the hole localization in CaO doped with Li and Na and the F-center in 
LiF as reported by Dovesi {\it et al.} \cite{Dovesi}; dimerization in C$_{4N+2}$ 
carbon rings \cite{Torelli}; the structure of the Al defect in $\alpha$-quartz 
\cite{Pacchioni,Stokbro,Magagnini}; exciton self-trappping in presence of 
thermal disorder \cite{ICDIM2000,Gavartin}; the phase diagram of crystalline Pu 
(related to a localization of $f$-type atomic orbitals) \cite{Pu1,Pu2,Pu3}.

On the other hand, the use of the
Hartree-Fock based approaches is also limited since neglecting electron 
correlation favours localized states \cite{Fulde}. One should also note
that frequently used approaches based on tight binding methods do not 
fully account for the electron kinetic energy, and thus are also intrinsically 
biased towards localized solutions. 
As seen from these arguments, the DFT approach in KS formalism is unable to 
reliably resolve between localized and delocalized situations in systems 
where both type of states are possible. A more adequate many-electron theory
for the self-trapping problem must go beyond the KS methods. Different forms of 
the density functionals free of the self-interaction error should be examined.  

\begin{acknowledgments}
The authors gratefully acknowledge funding by the Leverhulme Trust and NATO 
grant CRG.974075. Calculations have been performed on the CRAY T3E 
supercomputer facility provided by the Materials Chemistry Consortium, UK, 
and on the Bentham computer at the HiPerSpace Centre at the University College 
London. JLG is grateful to A. Sokol for many stimulating discussions, and
to E. Kotomin, E. Heifets, L. Kantorovich, G. Pacchioni, and C. Bird for 
their useful comments on the manuscript.
\end{acknowledgments}

\bibliography{vk_center}

\begin{thebibliography}{60}
\bibitem[{not()}]{note1}
\bibinfo{note}{In the supercell of 32 atoms, the 6 most remote atoms from the
  exciton were constrained. In the superecell of 64 atoms, atoms located in the
  (100) plane most remote from and parallel to the (100) plane of the exciton
  were constrained during the geometry optimisation. All other atoms were
  relaxed. In the supercells containing 108 and 144 atoms, all the atoms on
  which the force exceeded 0.04 eV$\AA^{-1}$ were included in the relaxation.}
\end{thebibliography}

\end{document}